\documentclass[aps,twocolumn]{revtex4}
\usepackage{graphicx}
\usepackage{bm}
\pdfoutput=1
\usepackage{braket}
\usepackage{pifont}
\usepackage{amsmath}
\usepackage{amsfonts}
\usepackage{amssymb}
\usepackage{graphicx}
\usepackage[FIGBOTCAP]{subfigure}
\usepackage{color}
\usepackage{tensor}
\graphicspath{{figs/}}
\usepackage{natbib}
\newcommand {\mrm}[1] {\mathrm{#1}}

\begin{document}

\title{Quantum Repeaters Using Continuous Variable Teleportation} 

\author{Josephine Dias}
\email{josephine.dias@uqconnect.edu.au}
\author{T.C. Ralph}

\affiliation{Centre for Quantum Computation and Communication Technology$,$ School of Mathematics and Physics$,$ University of Queensland$,$ Brisbane$,$ Queensland 4072$,$ Australia}

\date{\today}

\begin{abstract}
Quantum optical states are fragile and can become corrupted when passed through a lossy communication channel. Unlike for classical signals, optical amplifiers cannot be used to recover quantum signals. Quantum repeaters have been proposed as a way of reducing errors and hence increasing the range of quantum communications. Current protocols target specific discrete encodings, for example quantum bits encoded on the polarization of single photons. We introduce a more general approach that can reduce the effect of loss on any quantum optical encoding, including those based on continuous variables such as the field amplitudes. We show that in principle the protocol incurs a resource cost that scales polynomially with distance. We analyse the simplest implementation and find that whilst its range is limited it can still achieve useful improvements in the distance over which quantum entanglement of field amplitudes can be distributed. 
\end{abstract}

\maketitle

\section{Introduction}
Quantum communication enables various cryptographic protocols that outperform their classical counterparts including Quantum Key Distribution (QKD), with its promise of absolutely secure transmission of information \cite{gisin2002quantum}. The use of quantum optical systems as information carriers is currently the only practical approach to quantum communication  \cite{bachor2004guide}. Never-the-less, one of the biggest challenges facing the realisation of long distance quantum communication is optical loss due to fibre or free-space attenuation. One proposed method to enable long distance transmission of quantum states is the quantum repeater \cite{briegel1998quantum}. In this model, a lossy quantum channel is segmented into smaller, more manageable attenuation lengths along which entanglement is distributed and then purified. Entanglement swapping operations are then performed resulting in entanglement being held between both ends of the quantum channel. 

There have been a number of proposals for quantum repeaters that work on discrete variable quantum systems such as the polarization of single photons, and some elements of these have been implemented experimentally \cite{sangouard2011quantum}. However, quantum communication protocols can also be implemented using quantum continuous variables \cite{weedbrook2012gaussian}. Intriguingly, continuous variable entanglement swapping protocols can swap any optical entanglement, whether over discrete or continuous variables \cite{takeda2013deterministic,polkinghorne1999continuous},  and protocols for the purification of continuous variable entanglement have been developed \cite{xiang2010heralded, ulanov2015undoing}. This suggests that continuous variable (CV) quantum repeaters may be possible, and more versatile than discrete variable devices. Indeed, one might expect that a CV quantum repeater would correct errors on quantum information sent through optical modes, independently of how it was encoded. However, significant challenges exist to realising such a device.

To date, a complete quantum repeater protocol for continuous variables has not been described, although evidence that CV quantum repeaters can increase transmission distances has been presented \cite{campbell2013continuous} and hybrid protocols combining continuous and discrete states have been proposed \cite{van2008quantum}. It is known that regenerative stations containing only Gaussian elements cannot act as CV quantum repeaters \cite{namiki2014gaussian}.  

In this paper, we outline an architecture for a CV quantum repeater that relies on concatenated error correction protocols consisting of continuous variable teleportation \cite{braunstein1998teleportation} and entanglement distillation via noiseless linear amplification \cite{ralph2009nondeterministic}.  The paper is arranged in the following way. In the next section we review the continuous variable error correction protocol from Ref~\cite{ralph2011quantum}. In section \ref{sec:repeater}, we will describe how the error correction can be concatenated in such a way that the same effective transmission coefficient is maintained even though the physical channel is growing in length. We show that ideally the overhead for this concatenation is polynomial in the length of the channel. In section \ref{sec:1QS}, we numerically evaluate the performance of the CV quantum repeater assuming the simplest implementation of noiseless linear amplification. We find the range is limited under these conditions, however the device can still distribute continuous variable Einstein, Podolsky, Rosen (EPR) entanglement over significant distances. 


\section{The Error Correction Protocol}
We begin by reviewing the error correction protocol for continuous-variable states described in Ref~\cite{ralph2011quantum}. This technique for quantum error correction is effective against Gaussian noise induced by loss and proceeds by distilling EPR entanglement and using this entanglement for teleportation.
\begin{figure}
\centering
\subfigure[]{\label{fig:loss-channel}\includegraphics[width=\linewidth]{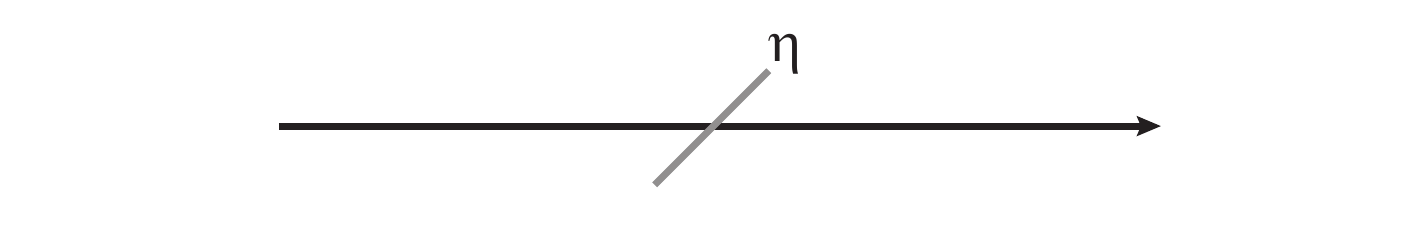}}
\subfigure[]{\label{fig:error-correction1}\includegraphics[width=\linewidth]{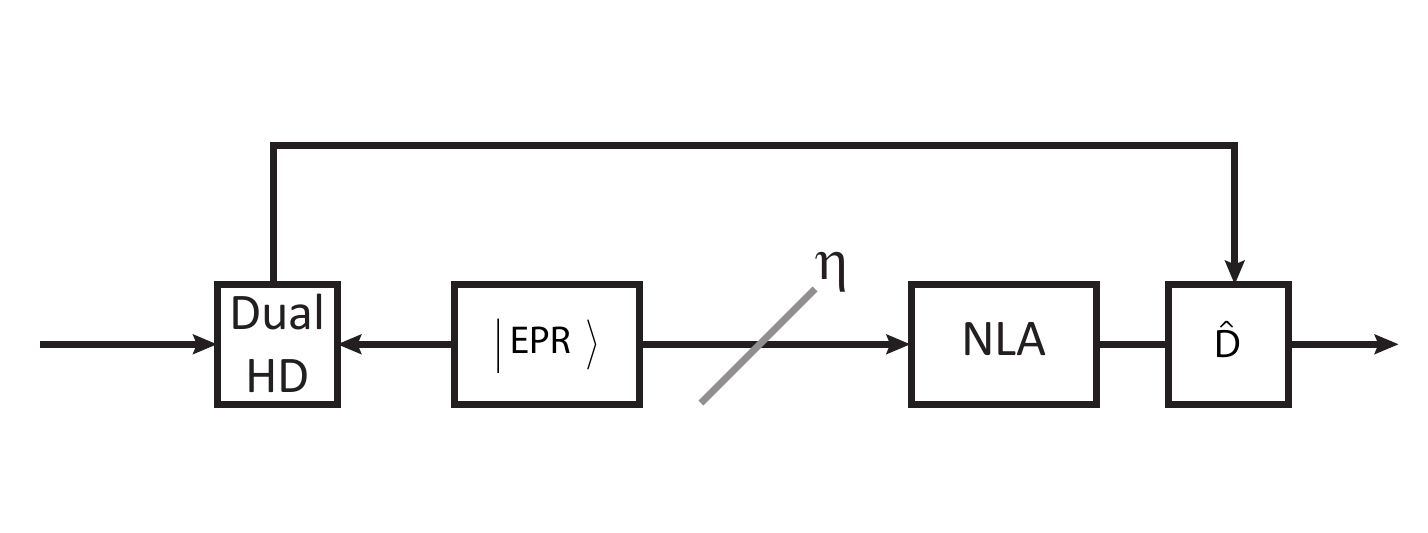}}
\caption{\subref{fig:loss-channel} A lossy channel of transmission \(\eta\). This channel takes an input coherent state \(\ket{\alpha}\) to output state \(\ket{\sqrt{\eta} \alpha}\).  \subref{fig:error-correction1} Protocol for quantum error correction against loss from Ref~\cite{ralph2011quantum}. EPR entanglement is distributed through a lossy channel of transmission \(\eta\). Noiseless linear amplification is then performed to distill the entanglement which is used for teleportation.}
\label{fig:error-correction}
\end{figure}
The aim of the protocol is to improve the effective transmission of any quantum state passing through a lossy channel (Fig.1(a)). The protocol is pictured in Fig.\ref{fig:error-correction1} where an EPR (or two mode squeezed) state is distributed through the lossy channel. Distillation is achieved via noiseless linear amplification (NLA) \cite{ralph2009nondeterministic} which is non-deterministic but heralded. When successful, the effect of the NLA on the entanglement is to produce an EPR state of higher purity (for a given entanglement strength) than achievable via direct transmission through the channel \cite{ulanov2015undoing}. After successful operation of the NLA, the distilled entanglement is used for teleportation: the input signal and the arm of the entangled state that did not pass through the loss are mixed on a 50:50 beamsplitter and conjugate quadratures are detected on each output mode via homodyne detection (also known as dual homodyne detection); the results of the measurement are sent via a classical channel to the receiver; and amplitude and phase modulation proportional to the measurement result are performed to displace the arm of the entanglement that passed through the loss and the NLA, producing the output mode.

For an input coherent state \(\ket{\alpha}\), the action of the lossy channel causes the  transformation:
\begin{equation}
\ket{\alpha}\to\ket{\sqrt{\eta}\alpha}
\label{eq:transform loss}
\end{equation}
where \(\eta\) is the transmission of the channel. In contrast, if the input coherent state is instead teleported using the distilled EPR state using the gain tuning protocol \cite{polkinghorne1999continuous, takeda2013deterministic} we obtain the transformation:
\begin{equation}
\ket{\alpha}\to\ket{g\sqrt{\eta}\chi\alpha}
\label{eq:transform}
\end{equation}
where \(g\) is the gain of the NLA, and \(\chi\) is the strength of the entanglement. By controlling the gain of the NLA, the effective transmission of the channel can be controlled. In particular, we will be interested in the case where \(g\) is chosen to be \(\frac{1}{\eta^{1/4}\chi}\) and the output \eqref{eq:transform} of the protocol is \(\ket{\eta^{1/4}\alpha}\). That is, the channel of transmission \(\eta\) has been error corrected to an effective transmission of \(\eta_{\mrm{eff}}=\sqrt{\eta}\). 

It should be noted that the transformation \eqref{eq:transform} is only exactly achieved when the NLA operates in an unphysical asymptotic limit. When implemented with linear optics, the NLA can be constructed from an array of \(N\) modified quantum scissors devices (a single quantum scissor is shown in Fig.~\ref{fig:QS1_NLA}) \cite{pegg1998optical}. The input state is split evenly among the \(N\) quantum scissors devices and the state is truncated in the photon number basis to order \(N\). This inevitably limits the fidelity between the target and output states of the NLA and hence compromises the operation of the error correction unless \(N\gg1\). In addition, the success probability of the NLA decreases exponentially with the number of quantum scissors -- thus imposing a significant resource cost on achieving high fidelity. 
Never-the-less, as was shown in Ref~\cite{ralph2011quantum}, this protocol can still be effective at correcting errors induced by loss on field states in the high loss regime.

\section{ The CV Quantum Repeater \label{sec:repeater}}
We now construct a quantum repeater by concatenating these error correction protocols in such a way that the effective transmission of the quantum channel is constant with distance. 

\begin{figure}
\centering
\subfigure[]{\label{fig:Concatenate1}\includegraphics[width=\linewidth]{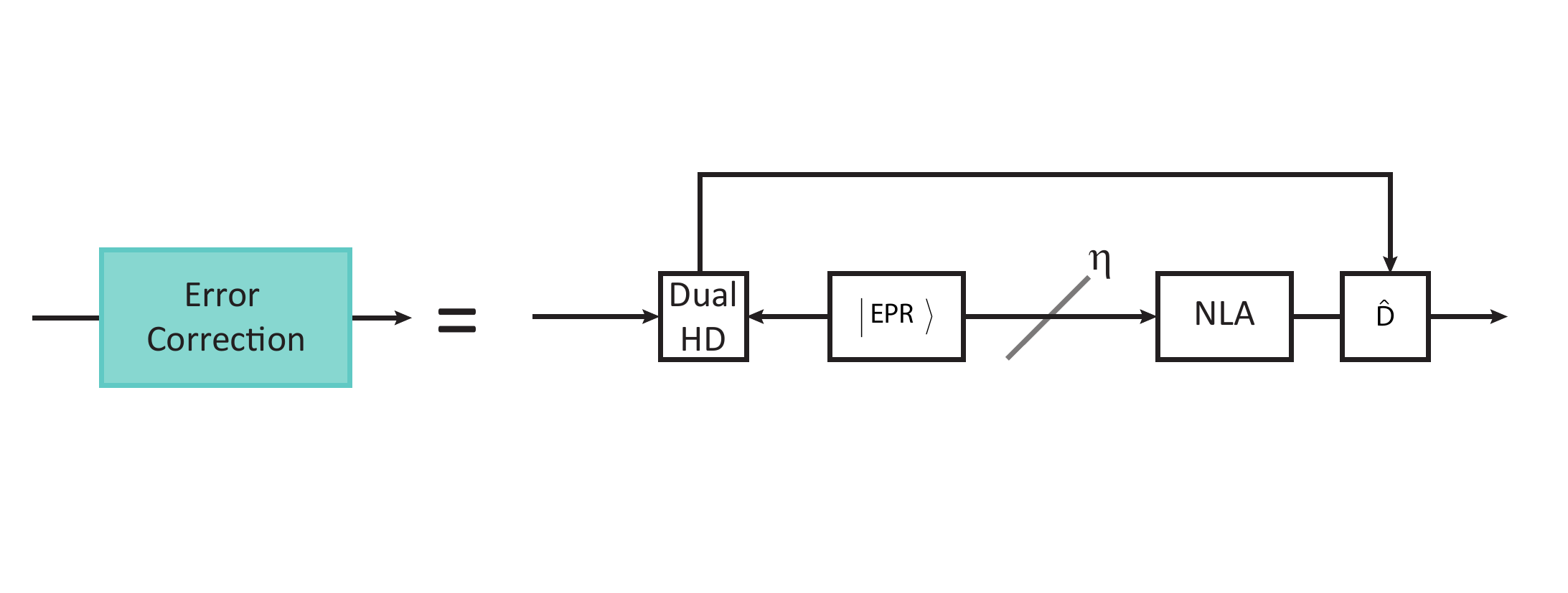}}
\subfigure[]{\label{fig:Concatenate2}\includegraphics[width=\linewidth]{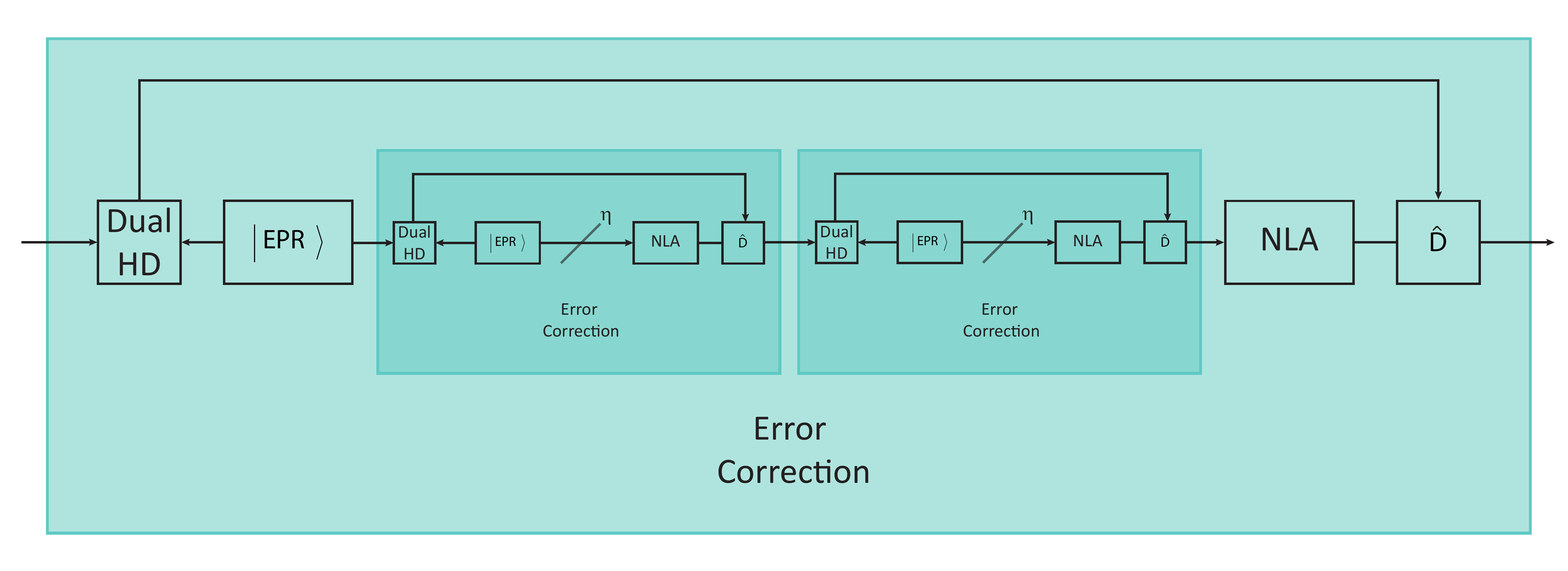}}
\subfigure[]{\label{fig:Concatenate3}\includegraphics[width=\linewidth]{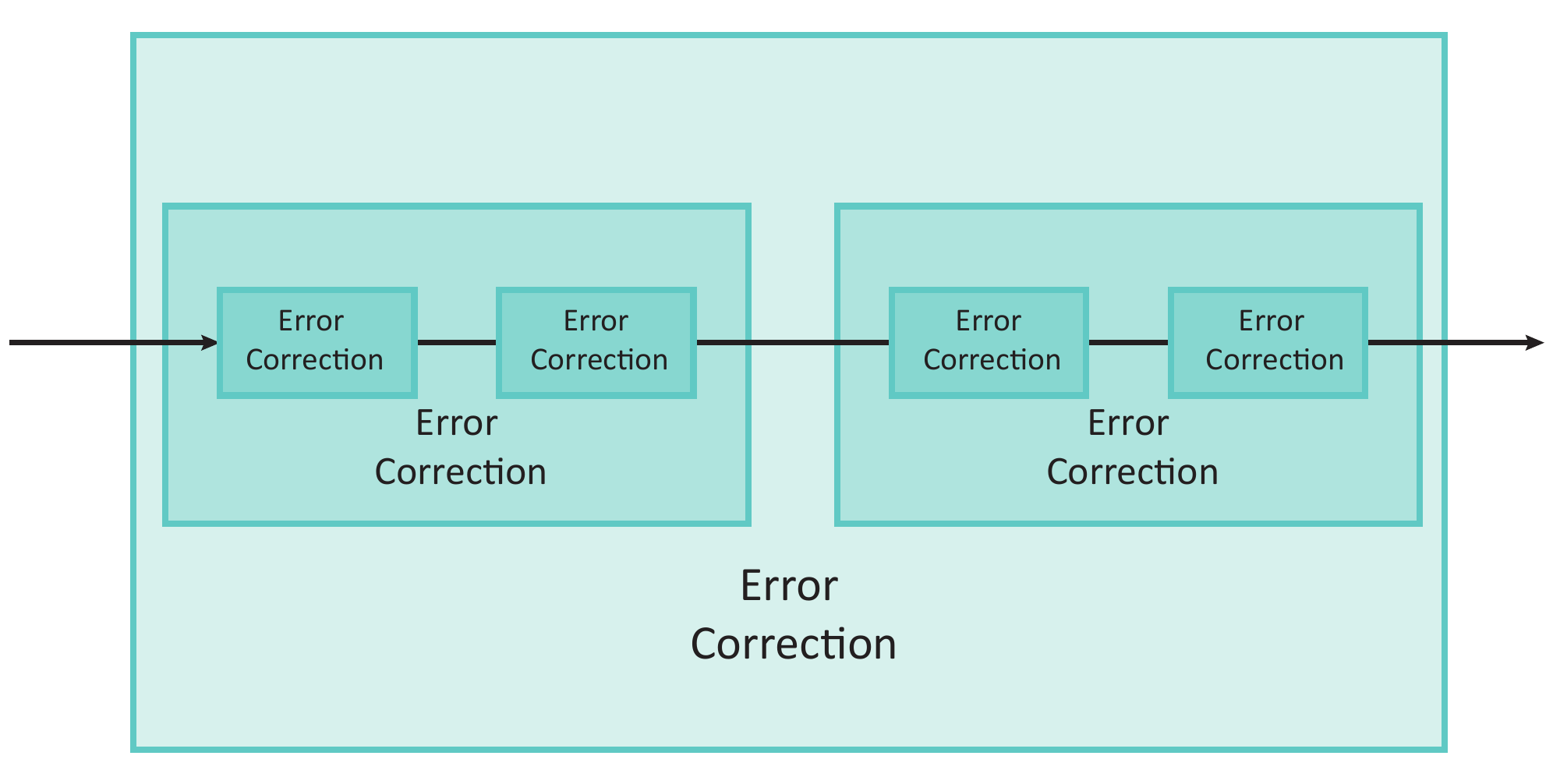}}
\caption{Structure of the quantum repeater for continuous variable states.  \subref{fig:Concatenate1} $M=1$. A single iteration of the error correction protocol to correct a lossy channel of transmission \(\eta\).  By choosing the gain of the NLA to be \(\frac{1}{\eta^{1/4}\chi}\) we can increase the transmission of the channel to be effectively \(\sqrt{\eta}\).  \subref{fig:Concatenate2} $M=2$. Explicit structure of two links of the repeater for double the distance (initial transmission \(\eta^2\).) Two error correction boxes are nested inside a larger error correction which represents the replacement of the physical lossy channel within Fig.~\ref{fig:error-correction1}. The two nested protocols error correct to produce effective channel transmission of \(\eta\) and the larger protocol produce overall effective transmission of \(\sqrt{\eta}\). \subref{fig:Concatenate3} $M=4$. Another doubling of the distance results in a channel of transmission \(\eta^4\) which is corrected by nesting error correction protocols in each other as shown. In this diagram we have used the schematic boxes; each box labelled ``error correction" corresponds to the error correction protocol depicted in Fig.~\ref{fig:error-correction1} and \ref{fig:Concatenate1}. At the base level, the channel is broken into four segments each of transmission \(\eta\) and error correction is performed on each segment (taking individual segments to effective transmission \(\sqrt{\eta}\).) Two error corrected segments effectively produce a channel of transmission \(\eta\), which are then nested in larger error correction protocol. Overall, this \(\eta^4\) channel has been error corrected to effective transmission \(\sqrt{\eta}\).}
\label{fig:Concatenation}
\end{figure}

The repeater is depicted in Figure~\ref{fig:Concatenation} where we show the structure of the protocol for increasing distance. Each individual error correction box (representing the protocol shown in Fig.~\ref{fig:error-correction1}) takes the initial transmission of the channel \(\eta\) to an effective transmission \(\sqrt{\eta}\) by using the gain condition:
\begin{equation}
g=\frac{1}{\eta^{1/4} \chi}
\label{eq:gain} 
\end{equation}

One iteration of the protocol, as shown in Fig.~\ref{fig:Concatenate1}, corrects a channel of transmission \(\eta\) to \(\sqrt{\eta}\). To preserve this transmission over double the distance (initial transmission of \(\eta^2\)), we use the protocol shown in Fig.~\ref{fig:Concatenate2}. Two nested protocols take the direct channel transmission \(\eta^2\) to effective transmission \({\eta}\). These nested error correction protocols take the position of the loss channel in shown in Fig.~\ref{fig:Concatenate1}. The larger protocol then corrects this to \(\sqrt{\eta}\). 

To preserve this transmission \(\sqrt{\eta}\) over another doubling of distance, another two links of the repeater are necessary as in Fig.~\ref{fig:Concatenate3}. The four base level error correction protocols work to correct a channel of transmission \(\eta^4\) to  \(\eta^2\). These base level protocols are nested within two higher level error correction protocols allowing the effective transmission \(\eta^2\) to be further corrected to \(\eta\). The last and highest level of error correction then produces a channel of effective transmission \(\sqrt{\eta}\).  Concatenation proceeds in this way for increasing distance where a channel of transmission \(\eta^{2^k}\) requires \(k\) levels of concatenation. 

When run in series, two error correction protocols may operate their NLAs independently and simultaneously. Throughout this paper we implicitly assume that high quality quantum memories are available that can store quantum states without loss of fidelity till the synchronising signals arrive from the various NLAs. 

At the first level of concatenation, the individual error correction procedures need to herald successful operation before error correction at the next level of concatenation can proceed. 
Therefore, if \(P\) is the success probability for one iteration of the error correction protocol Fig.~\ref{fig:Concatenate1}, then the entire protocol in Fig.~\ref{fig:Concatenate2}  operates with success probability \(P^2\).  Similarly, the success probability for the protocol in Fig.~\ref{fig:Concatenate3} is \(P^3\). Whilst the probability of success is dropping exponentially with the number of concatenations, the distance doubles. In general, we have:
\begin{equation}
P_M = P^{\log_2 \left(2 M\right)} =  \left(2 M\right)^{\log_2 P}
\label{eq:Pscale}
\end{equation}
where \(M\) is the number of links of the quantum repeater, and thus we obtain a polynomial scaling of success probability with distance.  

We can estimate $P$ in the following way. For a particular gain, $g$, the NLA has a probability of success $P_g \approx 1/(g+1)^{2N}$ (see the Appendix). Inserting the gain condition Eq. \ref{eq:gain} and assuming $g\gg1$ we obtain:
\begin{equation}
P \approx ( \eta\chi^4 )^{N/2}
\end{equation}
\label{prob}
To evaluate the efficiency of the device we can compare the probability of successfully sending a single photon through the error corrected channel, $\sqrt{\eta} P_M$, to the probability of successfully sending a single photon through the bare channel, $\eta^M$. In this way we can obtain the desirable condition:
\begin{equation}
P_M \approx (2 M)^{{{N}\over{2}} Log_2(\chi^4 \eta)} > \eta^{M-{{1}\over{2}}}.
\end{equation}
Because of the exponential scaling of the bare channel it is clear that there will always be an $M$ at which the quantum repeater will be more efficient than the bare channel, however whether that break even point occurs whilst $P_M$ still has a practical value depends on the choice of parameters (and what one considers a practical value). As an example if we pick $\eta = 0.04$, $\chi = .9$ and $N = 3$ we find the break-even point is around $M = 8$.
For these parameters, we obtain $P_M \approx 3 \times 10^{-10} >> \eta^{M-{{1}\over{2}}} \approx 3 \times 10^{-11}$.

The limitation of this efficiency scaling argument is that it ignores potential truncation noise that might build with each level of concatenation. We now examine an example of this with the simple case of the repeater protocol where the NLA is implemented with a single quantum scissor. 

\begin{figure}
\centering
\includegraphics[width=0.7\linewidth]{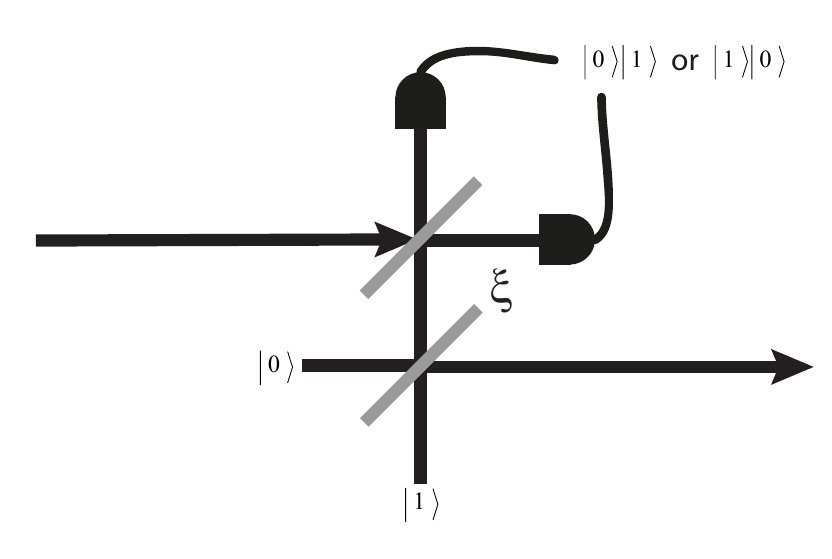}
\caption{Structure of the NLA \cite{ralph2009nondeterministic} when it consists of a single modified quantum scissor device \cite{pegg1998optical}. The NLA is successful when the single photon detectors register one photon at one detector and zero photons at the other. The beam splitter ratio \(\xi\) is related to the gain of the NLA by \(g = \sqrt{\left(1-\xi\right) /\xi}\). \label{fig:QS1_NLA}}
\end{figure}

 \section{Operation with a single quantum scissor\label{sec:1QS}}
 
Of immediate practical interest is the performance of the system in the simplest case where the NLA is constructed from a single quantum scissor. For such a situation we expect truncation noise to be significant. Hence we now examine operation of the repeater protocols shown in Figures~\ref{fig:Concatenate1} and \ref{fig:Concatenate2} for the case where the NLA consists of the single quantum scissor device shown in Fig.~\ref{fig:QS1_NLA}. This device performs the transformation 
\begin{equation}
\hat{T}_1(\alpha\ket{0} +\beta\ket{1}) = \sqrt{\frac{1}{g^2+1}}(\alpha\ket{0} + g\beta\ket{1})
\label{eq:single_qs}
\end{equation}
with all higher order terms truncated. The effect of this truncation on the protocol is to increase the variance of the output state above the quantum noise limit level expected for a coherent state. We refer to this as `truncation noise.'

As a first figure of merit for our protocol we ask if the level of truncation noise introduced is low enough to allow entanglement distribution through the channel. A sufficient condition for entanglement distribution is that the excess noise \(\delta\) is bounded by \(\delta<2\eta\) \cite{namiki2004practical}. When the noise added is above this bound, the state may be separable and so not useful for quantum communication. We have calculated the variance of the output state of the protocols  in Figures~\ref{fig:Concatenate1} and \ref{fig:Concatenate2}. This is dependent on the entanglement strength of the two mode squeezed state, \(\chi\), and the transmission of the channel between nodes, \(\eta\). We note that because the truncation noise is non-Gaussian, \(\delta<2 \eta\) is not a necessary condition for entanglement breaking - a point we will return to later.

\begin{figure}
\centering
\includegraphics[width=0.7\linewidth]{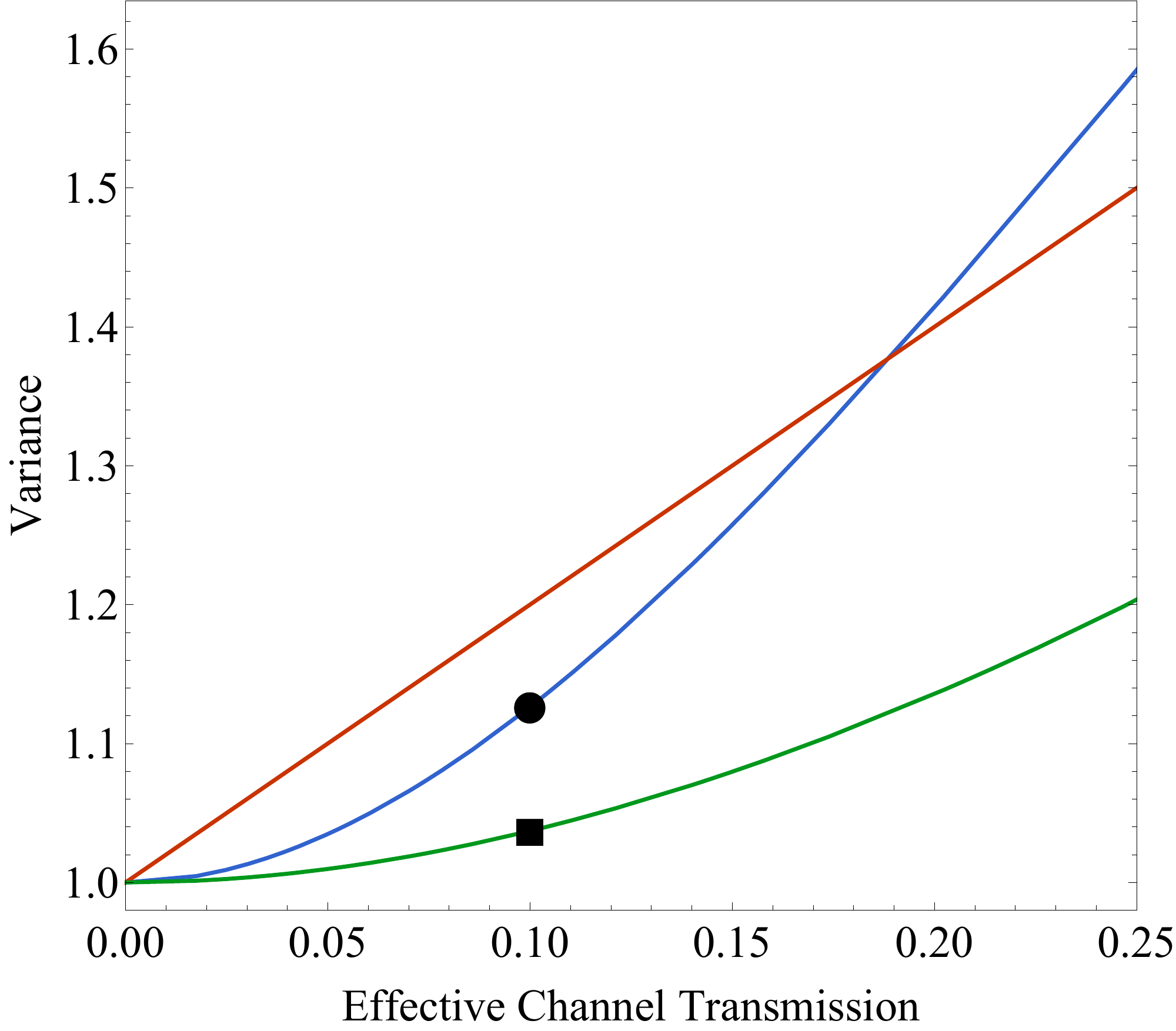}
\caption{Variance of the protocol in Fig.~\ref{fig:Concatenate1} plotted against the effective transmission of the channel. This protocol takes a channel of direct transmission \(\eta\) to one of effective transmission \(\sqrt{\eta}\) using \(\chi = 0.7\)  (blue) and \(\chi=0.1\) (green). Also shown is the entanglement breaking bound (red). The point \textcolor[rgb]{0,0,0}{\ding{108}} is achieved with a success probability of \(P=0.06\) and the point \textcolor[rgb]{0,0,0}{\ding{110}} is achieved with \(P= 0.001\). \label{fig:VarOne}}
\end{figure}

\begin{figure}
\centering
\includegraphics[width=0.7\linewidth]{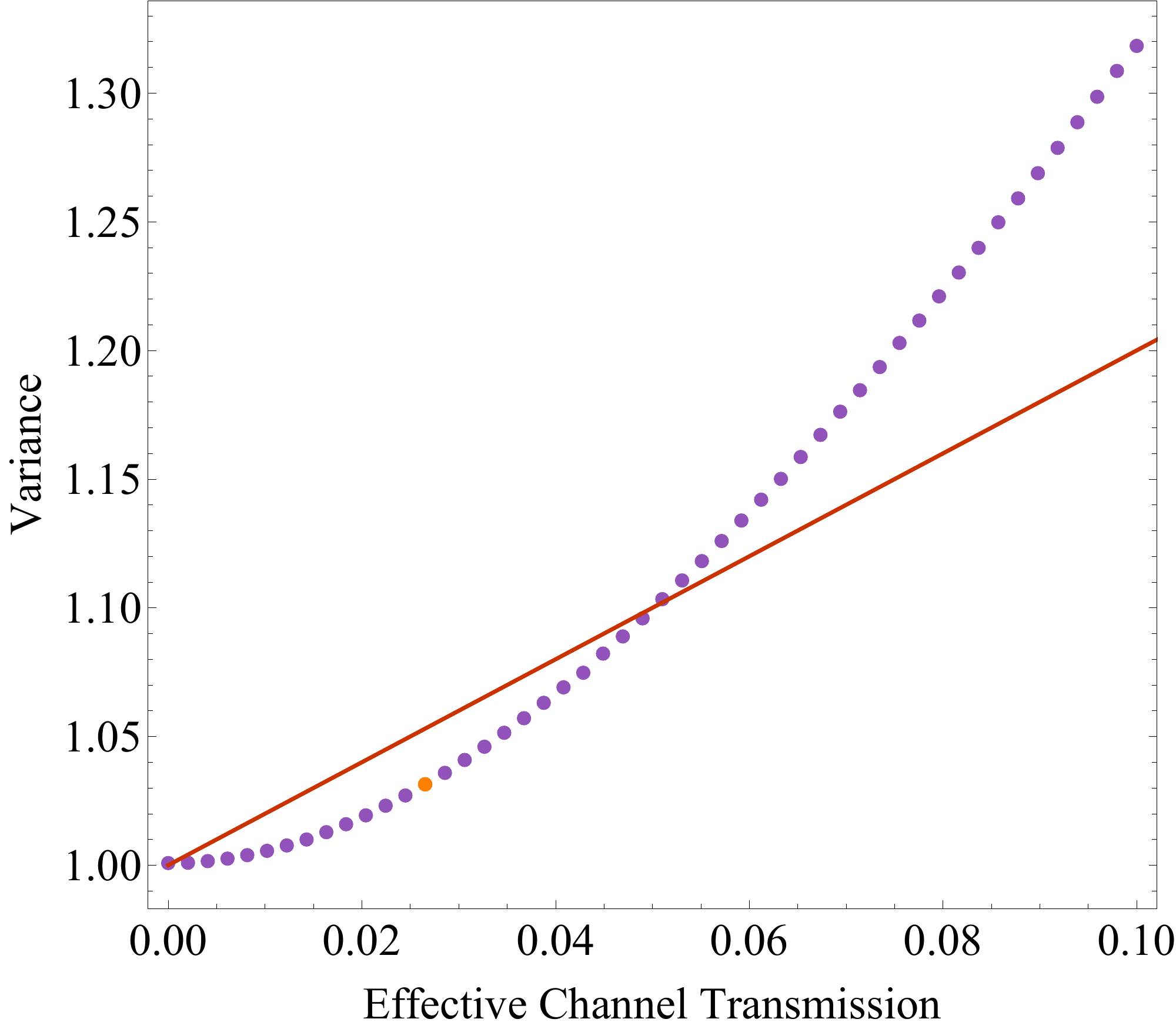}
\caption{Variance of the protocol for two links of the repeater shown in Fig.~\ref{fig:Concatenate2}(purple) plotted against the effective transmission of the channel. This protocol takes a channel of direct transmission \(\eta^2\) to one of effective transmission \(\sqrt{\eta}\). The nested error correction protocols use \(\chi=0.01\) and the larger protocol used \(\chi=0.7\). Also shown is the entanglement breaking bound (red). The point \textcolor[rgb]{1,0.5,0}{\ding{108}}  is achieved with success probability \(4\times10^{-8}\). \label{fig:VarConc} }
\end{figure}

The results contained in Fig.~\ref{fig:VarOne}  show the variance of one link of the repeater.  These calculations are detailed in the Appendix and assume ideal detectors, and single photon and EPR sources.  This protocol preserves the effective transmission of a channel \(\sqrt{\eta}\) over double the actual distance \(\eta\). We observe here a significant difference in outcome when using a high strength two mode squeezed state (\(\chi=0.7\)) to that of a weakly entangled state (\(\chi=0.1\)). When using a weakly entangled state, the excess noise produced is sufficiently small such that the channel is entanglement preserving for any transmission \(\eta\). While this outcome is favourable in terms of excess noise produced, this requires using a higher gain in the NLA, and therefore comes with a decrease in success probability. To illustrate this effect, the two points in Fig.~\ref{fig:VarOne} both take an initial loss channel of 1\% and improve it to  effectively 10\%; using \(\chi=0.7\) this can be done with success probability \(P=0.06\) and with \(\chi=0.1\) the success probability is \(P=0.001\).


Fig.~\ref{fig:VarConc} shows the variance of the concatenated protocol shown in Fig.~\ref{fig:Concatenate2} (achieving an effective transmission \(\sqrt{\eta}\) from an initial \(\eta^2\)). This protocol is also capable of operating under the entanglement breaking bound albeit in a high loss regime only. These results were obtained using very weakly entangled EPR states (\(\chi=0.01\)) in the nested error correction protocols. Operating the concatenated protocol of Fig.~\ref{fig:Concatenate2} in this way ensures the nested error correction protocols contribute minimal excess noise. Then using a higher strength EPR state (\(\chi=0.7\)) in the larger error correction protocol ensures the gain of the final NLA is reduced and thus avoids amplifying the truncation noise produced by the nested protocols.  This represents a tradeoff in the operation of this concatenated protocol, where the desired outcome of excess noise being within the entanglement preserving regime is only achieved with restrictions on the parameters \(\chi\) and \(\eta\). Employing more quantum scissors in the NLAs would reduce noise and enable this protocol to be useful at higher effective transmissions. However, the cost is an exponential decrease in probability of success with increasing numbers of quantum scissors.

\begin{figure}
\centering
\includegraphics[width=0.7\linewidth]{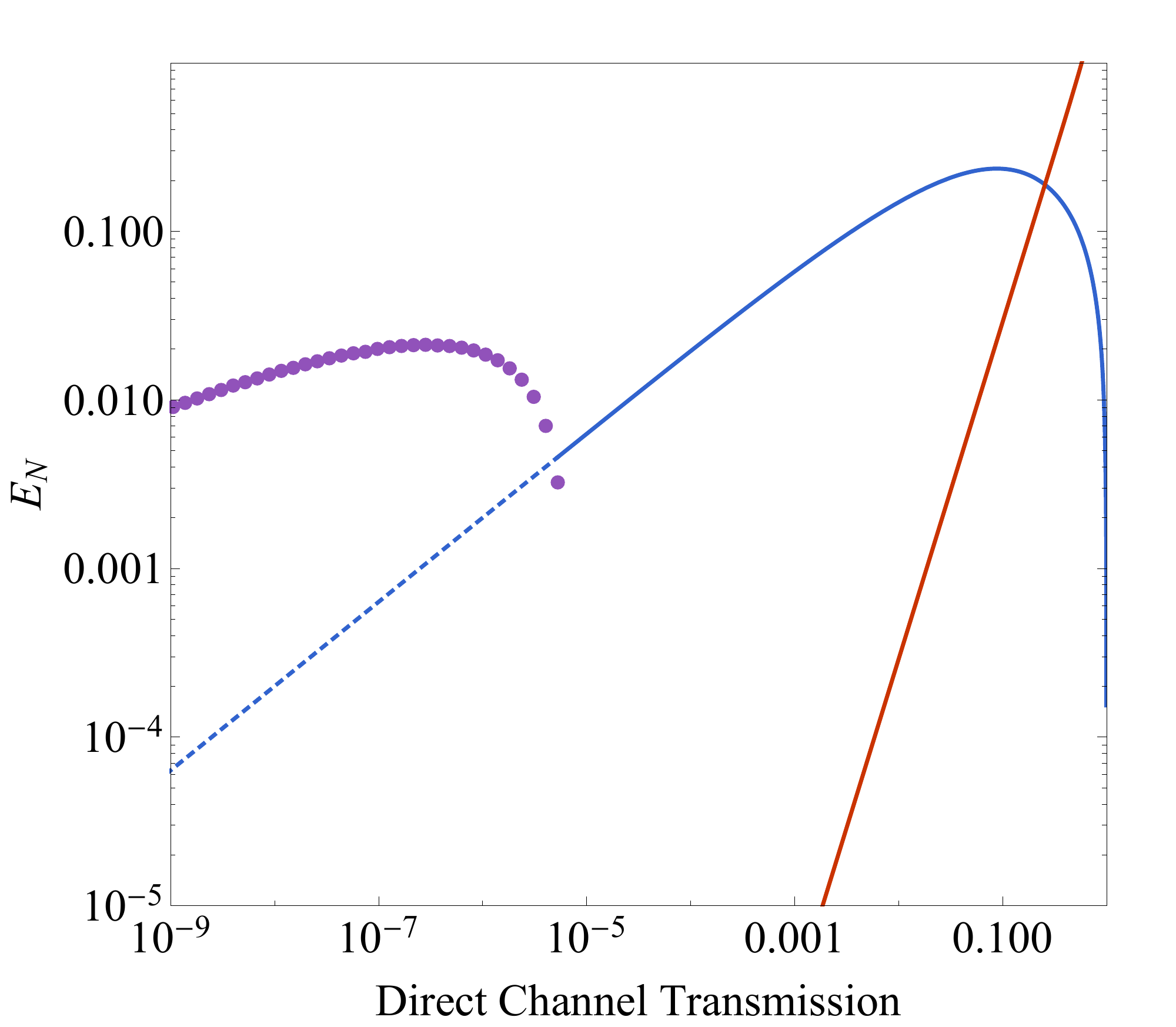}
\caption{Logarithmic negativity of the protocol plotted against direct channel transmission. One iteration of the repeater shown in Fig.~\ref{fig:Concatenate1} (blue line) using \(\chi=0.01\). The concatenated protocol shown in Fig.~\ref{fig:Concatenate2} (purple), using \(\chi=0.01\) for the nested error correction protocols and \(\chi=0.7\) for the higher level protocol. The deterministic entanglement limit with direct transmission (red). The region abover the red line is where the protocol outperforms the bare channel. \label{fig:En}}
\end{figure}

Now that we have shown that these protocols are useful for quantum commmunication, we can also examine how well this channel reduces entanglement degredation caused by loss by using the logarithmic negativity \cite{vidal2002computable}. This gives us an unambiguous measure of improvement of the bare channel achieved by the protocol. Because we assume Gaussian noise, this represents a lower bound on the protocol's performance. In Figure~\ref{fig:En} we compare the logarithmic negativity achieved by the protocols shown in Fig.~\ref{fig:Concatenate1} and Fig.~\ref{fig:Concatenate2} to that of an EPR state distributed through the same loss in the limit of infinite squeezing. This is given by \(E_N^\infty = -\log_2 \frac{1-\eta}{1+\eta}\) \cite{ulanov2015undoing}. We demonstrate here a region where our device achieves a significant improvement over that which would be achieved with a perfect EPR state using direct transmission. 

The device we present here is unfortunately limited in its operational efficiency by low success probabilities. While an initial loss channel of 1\% can be improved to  effectively 10\% with a success probability of 0.06, the concatenated protocol (Fig.~\ref{fig:Concatenate2}) takes an initial loss channel of \( 5\times10^{-5} \%\) and increase it to effectively \(  2 \% \) with a probability of success of \( 4\times 10^{-8}\). As stated earlier, these calculations were obtained with the minimum number of quantum scissors employed in the NLAs.

Another problematic aspect of this protocol is the highly restricted regime in which its useful to transfer quantum entanglement. That is, the very high loss regime. In cases where it is necessary to operate the device at higher effective transmissions, one must employ more quantum scissors in the NLAs and this comes with the unfortunate cost of a reduction in the probability of success. This signifies the most prominent limitation in using this repeater; that is less excess noise comes at the expense of the probability of success. 

While the results presented in this paper were generated with NLAs containing a single quantum scissor, it may be worthwhile to consider the outcomes when the devices consist of two quantum scissors. In this case, the transformation \eqref{eq:single_qs} would be replaced by
\begin{multline}
\hat{T}_2\left(\alpha\ket{0} +\beta\ket{1}+\gamma\ket{2}\right) =
\\ {\frac{1}{g^2+1}}\left(\alpha\ket{0} + g\beta\ket{1}+ \frac{1}{2}g^2\gamma\ket{2}\right)
\end{multline}
which now keeps the \(\ket{2}\) terms in the state and truncates higher order terms. We can expect a reduction in excess noise when two quantum scissors are used which is especially important when the error correction protocols are concatenated as in Fig.~\ref{fig:Concatenate2} and \ref{fig:Concatenate3}. Additionally, adding a single extra quantum scissor results in a minimal decrease in success probability. 

Results with the protocols in Fig.~\ref{fig:Concatenate2} and \ref{fig:Concatenate3} using two or more quantum scissors would be of significant interest. However, modelling these protocols is computationally intensive given the complex design of the concatenated error correction. Ideally, one would like to model these channels as effective Gaussian channels where the state truncation step introduces some Gaussian noise. However, when this is done, the results do not agree with that of the exact output states due to the NLA performing a non-Gaussian operation. Indeed, typically the performance is inferior to the exact result.  For this reason, the task of characterising the performance of the repeater for higher levels of concatenation and more quantum scissors remains extremely difficult.

\section{Discussion \label{sec:Discussion}}
In summary, we have proposed a method to concatenate error correction protocols to produce a quantum repeater that works with CV states. The error correction relies on continuous variable teleportation and entanglement distillation through noiseless linear amplification. Teleportation of CV states is advantageous because of its deterministic operation, but it also limits the channel transmission improvement achievable between input and output states. Importantly, the use of CV teleportation means the protocol will work on any field state and is therefore not limited to a particular optical encoding of quantum information. 

The repeater protocol we present here does have a polynomial efficiency scaling in the ideal case of the NLA. However, it is limited in practice due to the  trade-off between probability of successful operation and noise added from state truncation.  

This noise penalty due to state truncation is inevitable with the linear optics construction of the NLA. To this problem, we have shown that our protocol with a single quantum scissor in the NLA for $M=2$ links of the repeater (Fig~\ref{fig:Concatenate2}) can operate with sufficiently low added noise such that it can transmit entanglement of higher quality than the bare channel. 

 It is important to consider more generally under what conditions the repeater may be more efficient than direct transmission while simultaneously adding low enough noise so that the channel can be used to transmit entanglement. These conditions remain an open question because characterising the performance of our repeater for higher concatenation levels represents a computationally intensive task due to the structural complexity and the inability to model the device using Gaussian operations.

 As such, there remains significant room for improvement with this protocol. It remains an open question as to how the protocol may be amended to be useful at higher effective transmissions while maintaining (or improving) the probability of success. It is possible that non-linear optical techniques for implementing the NLA are needed to realise the full scaling potential of our device \cite{mcmahon2014optimal}.  

\section{Acknowledgements}
We thank Remi Blandino and Austin Lund for useful discussions. This research was funded by the Australian Research Council Centre of Excellence for Quantum Computation and Communication Technology (Project No. CE110001027).

\section{Appendix}
\subsection{The error correction protocol}
In this appendix we provide details on the error correction protocol of Fig.~\ref{fig:Concatenate1}. The continuous variable teleportation protocol uses shared entanglement of the form:
\begin{equation}
\ket{\chi}_{RB} = \sqrt{1-\chi^2} \sum_{n=0}^\infty \chi^n\ket{n}_R \ket{n}_B
\end{equation}

An arbitrary input state \(\ket{\psi}_A\) is mixed on a 50:50 beamsplitter with mode \(R\) and dual homodyne detection is performed. Here \(\beta_1\) is detected, where 
\begin{equation}
\beta_1 = X_-+iP_+
\end{equation}
with
\begin{align}
\hat{X}_- &= \hat{X}_A-\hat{X}_R 
\\
\hat{P}_+ &= \hat{P}_A+\hat{P}_R
\end{align} 
 This measurement projects onto the eigenstate \cite{hofmann2000fidelity}
 \begin{equation}
 \ket{\beta_1}_{AR} = \frac{1}{\sqrt{\pi}}\sum_{n=0}^\infty \hat{D}_A(\beta_1)\ket{n}_A\ket{n}_R 
 \end{equation}

The output state conditioned on the measurement result \(\beta_1\) is therefore
\begin{align}
\ket{\psi(\beta_1)} &= \tensor[_{AR}]{\braket{\beta_1|\psi}}{_{A}}  \ket{\chi}_{RB}
\\
& = \sqrt{\frac{1-\chi^2}{\pi}} \sum_{n, m} \chi^m \tensor[_A]{\bra{n}}{} \tensor[_R]{\bra{n}}{} \hat{D}^\dagger_A (\beta_1) \ket{\psi}_A  \ket{m}_R \ket{m}_B
\\
& = 
 \sqrt{\frac{1-\chi^2}{\pi}} \sum_n \chi^n \ket{n}_B \tensor[_A]{\bra{n}}{} \hat{D}_A (-\beta_1) \ket{\psi}_A
\end{align}
Where the measurement probability \(P(\beta_1)\) is given by \(\braket{\psi(\beta_1)|\psi(\beta_1)}\).

 For an input coherent state \(\ket{\psi}_A = \ket{\alpha}_A\)
 \begin{align}
 \tensor[_A]{\bra{n}}{} \hat{D}_A(-\beta_1)\ket{\alpha}_A &= \tensor[_A]{\braket{n|\alpha-\beta_1}}{_A} \nonumber
 \\
  &= e^{-|\alpha-\beta_1|^2/2} \frac{(\alpha-\beta_1)^n}{\sqrt{n!}}
 \end{align}
 
 \begin{align}
 \ket{\psi(\beta_1)} &= \sqrt{\frac{1-\chi^2}{\pi}}  e^{-|\alpha-\beta_1|^2/2} \underbrace{\sum_n \frac{(\chi(\alpha-\beta_1))^n}{\sqrt{n!}}\ket{n}_B}_{e^{|\chi(\alpha-\beta_1)|^2/2} \ket{\chi(\alpha-\beta_1)}}
\\
 &=\sqrt{\frac{1-\chi^2}{\pi}} e^{\frac{1}{2}|\alpha-\beta_1|^2(\chi^2-1)}\ket{\chi(\alpha-\beta_1)}
 \end{align}
 
 The state then passes through a lossy channel of transmission \(\eta\)
 \begin{equation}
 \ket{\psi(\beta_1)} =\sqrt{\frac{1-\chi^2}{\pi}} e^{\frac{1}{2}|\alpha-\beta_1|^2(\chi^2-1)}\ket{\sqrt{\eta}\chi(\alpha-\beta_1)}
 \end{equation}
 
 The action of the NLA with \(N\) quantum scissors can be described by the following operation:
\begin{equation}
\hat{T}_N = \hat{\Pi}_N g^{\hat{n}}
\end{equation}
where \(\hat{\Pi}_N\) is the truncation operator defined as:
\begin{equation}
\hat{\Pi}_N  = \left(\frac{1}{g^2+1}\right)^{\frac{N}{2}} \sum_{n =0}^N \frac{N!}{ (N-n)!N^n} \ket{n}\bra{n} 
\end{equation}
We are interested in the case where the NLA consists of a single quantum scissor, \(N=1\):
\begin{equation}
\hat{\Pi}_1 = \sqrt{\frac{1}{g^2+1}} \left(\ket{0}\bra{0}+\ket{1}\bra{1}\right)
\end{equation}
The state after action of the NLA is 
\begin{multline}
\ket{\psi(\beta_1)}=\sqrt{\frac{1-\chi^2}{1+g_1^2}} \frac{1}{\sqrt{\pi}} e^{\frac{1}{2}|\alpha-\beta_1|^2(\chi^2-1-\eta\chi^2)}
\\
\left(\ket{0}+g_1\sqrt{\eta}\chi(\alpha-\beta_1)\ket{1}\right)
\end{multline}

The last step in this protocol is a displacement by the measurement result \(\beta_1\) scaled by the gain of the NLA \(g_1\), the strength of entanglement \(\chi\) and transmission of the channel \(\eta\). 
The output state of the protocol is 
\begin{multline}
\ket{\psi_{out}(\beta_1)} = \sqrt{\frac{1-\chi^2}{1+g_1^2}} \frac{1}{\sqrt{\pi}} e^{\frac{1}{2}|\alpha-\beta_1|^2(\chi^2-1-\eta\chi^2)}
\\
\hat{D}(g\sqrt{\eta}\chi\beta_1) \left(\ket{0}+g_1\sqrt{\eta}\chi(\alpha-\beta_1)\ket{1}\right)
\label{eq:out1}
\end{multline}
The probability of success is 
\begin{align}
P &= \int \braket{\psi_{out}(\beta_1)|\psi_{out}(\beta_1)} \mathrm{d}^2 \beta  \label{eq:Pformula}
\\
P &= -\frac{\left(\chi ^2-1\right) \left(\chi ^2 \left(\eta  g^2+\eta -1\right)+1\right)}{\left(g^2+1\right) \left((\eta -1) \chi ^2+1\right)^2}
\end{align} 

\begin{widetext}
The variance is
\begin{align}
V &= \int \braket{\hat{X}^2} \mathrm{d}^2\beta -\left(\int \braket{\hat{X}}\mathrm{d}^2\beta\right)^2 \label{Vformula}
\\
V  &= \frac{\chi ^2 \left(\eta  \left(g^2+\chi ^2 \left(4 \eta  g^4+(\eta -1) g^2+\eta -2\right)+2\right)+\chi ^2-2\right)+1}{\left((\eta -1) \chi ^2+1\right) \left(\chi ^2 \left(\eta  g^2+\eta -1\right)+1\right)}
\end{align}

\subsection{Concatenated Protocol}

We now proceed to derive the exact output state of the concatenated error correction protocol shown in Fig.~\ref{fig:Concatenate2}. For this task, we begin by feeding the output state of the first protocol \eqref{eq:out1} into a second protocol. 

The first step is the joint measurement of \(\hat{X}\) and \(\hat{P}\) projecting the product state \(\ket{\psi_{out}(\beta_1)}_A\otimes\ket{\chi}_{RB}\) onto the state of mode \(B\)
\begin{equation}
\ket{\psi(\beta_1, \beta_2)} =\sqrt{\frac{1-\chi^2}{\pi}} \sum_n \chi^n \ket{n}_B \tensor[_A]{\bra{n}}{} \hat{D}_A (-\beta_2)\ket{\psi_{out}(\beta_1)}_A
\end{equation}
For the output state given in \eqref{eq:out1}, we have

\begin{equation}
\tensor[_A]{\bra{n}}{} \hat{D}_A (-\beta_2)\ket{\psi_{out}(\beta_1)}_A = \sqrt{\frac{1-\chi^2}{1+g_1^2}} \frac{1}{\sqrt{\pi}} e^{\frac{1}{2}|\alpha-\beta_1|^2(\chi^2-1-\eta\chi^2)}\tensor[_A]{\bra{n}}{} \hat{D}_A (-\beta_2) \hat{D}_A(g\sqrt{\eta}\chi\beta_1) \left(\ket{0}_A+g_1\sqrt{\eta}\chi(\alpha-\beta_1)\ket{1}_A\right)
\end{equation}
Using the following property of the displacement operator:
\begin{equation}
\hat{D}(\alpha)\hat{D}(\beta) = e^{(\alpha\beta^*-\alpha^*\beta)/2}\hat{D}(\alpha+\beta)
\label{displacement_operator_product}
\end{equation}
We may combine the displacements on mode \(A\) as:
\begin{equation}
\hat{D}_A (-\beta_2) \hat{D}_A(g\sqrt{\eta}\chi\beta_1)  = e^{g\sqrt{\eta}\chi(\beta_1\beta_2^*-\beta_1^*\beta_2)/2} \hat{D}_A(g\sqrt{\eta}\chi\beta_1-\beta_2)
\end{equation}
where \(\beta_1 \beta_2^*-\beta_1^*\beta_2\) is purely imaginary.
\begin{multline}
\tensor[_A]{\bra{n}}{} \hat{D}_A (-\beta_2)\ket{\psi_{out}(\beta_1)}_A = \sqrt{\frac{1-\chi^2}{1+g_1^2}} \frac{1}{\sqrt{\pi}} e^{g\sqrt{\eta}\chi(\beta_1\beta_2^*-\beta_1^*\beta_2)/2}  e^{\frac{1}{2}|\alpha-\beta_1|^2(\chi^2-1-\eta\chi^2)}
\\
\tensor[_A]{\bra{n}}{} \hat{D}_A(g\sqrt{\eta}\chi\beta_1-\beta_2) \left(\ket{0}_A+g_1\sqrt{\eta}\chi(\alpha-\beta_1)\ket{1}_A\right)
\end{multline}
\begin{equation}
\tensor[_A]{\bra{n}}{} \hat{D}_A(g\sqrt{\eta}\chi\beta_1-\beta_2)\ket{0}_A = \tensor[_A]{\braket{n|g\sqrt{\eta}\chi\beta_1-\beta_2}}{_A} = e^{-|g\sqrt{\eta}\chi\beta_1-\beta_2|^2/2} \frac{(g\sqrt{\eta}\chi\beta_1-\beta_2)^n}{\sqrt{n!}}
\end{equation}

\begin{equation}
\tensor[_A]{\bra{n}}{} \hat{D}_A(g\sqrt{\eta}\chi\beta_1-\beta_2)\ket{1}_A 
= e^{-|g\sqrt{\eta}\chi\beta_1-\beta_2|^2/2}\left( \sqrt{n}  \underbrace{\frac{(g\sqrt{\eta}\chi\beta_1-\beta_2)^{n-1}}{\sqrt{(n-1)!}}}_{n \geq 1}+(-g\sqrt{\eta}\chi\beta_1^*+\beta_2^*) \frac{(g\sqrt{\eta}\chi\beta_1-\beta_2)^n}{\sqrt{n!}}
\right)
\end{equation}
\begin{multline}
\tensor[_A]{\bra{n}}{} \hat{D}_A (-\beta_2)\ket{\psi_{out}(\beta_1)}_A = \sqrt{\frac{1-\chi^2}{1+g_1^2}} \frac{1}{\sqrt{\pi}} e^{g\sqrt{\eta}\chi(\beta_1\beta_2^*-\beta_1^*\beta_2)/2}  e^{\frac{1}{2}|\alpha-\beta_1|^2(\chi^2-1-\eta\chi^2)} e^{-|g\sqrt{\eta}\chi\beta_1-\beta_2|^2/2}
\\
\left( (1+g_1\sqrt{\eta}\chi(\alpha-\beta_1)(-g\sqrt{\eta}\chi\beta_1^*+\beta_2^*))\frac{(g\sqrt{\eta}\chi\beta_1-\beta_2)^n}{\sqrt{n!}} +g_1\sqrt{\eta}\chi(\alpha-\beta_1)\sqrt{n} \underbrace{ \frac{(g\sqrt{\eta}\chi\beta_1-\beta_2)^{n-1}}{\sqrt{(n-1)!}}}_{n \geq 1}\right)
\end{multline}
The state after measurement of \(\beta_2\) is
\begin{multline}
\ket{\psi(\beta_1, \beta_2)} =\frac{1-\chi^2}{\pi} \frac{1}{\sqrt{1+g_1^2}}  e^{g\sqrt{\eta}\chi(\beta_1\beta_2^*-\beta_1^*\beta_2)/2}  e^{\frac{1}{2}|\alpha-\beta_1|^2(\chi^2-1-\eta\chi^2)} e^{-|g\sqrt{\eta}\chi\beta_1-\beta_2|^2/2}
\\
\sum_{n=0}^\infty \chi^n  (1+g_1\sqrt{\eta}\chi(\alpha-\beta_1)(-g\sqrt{\eta}\chi\beta_1^*+\beta_2^*))\frac{(g\sqrt{\eta}\chi\beta_1-\beta_2)^n}{\sqrt{n!}}\ket{n}_B +\sum_{n=1}^\infty \chi^n g_1\sqrt{\eta}\chi(\alpha-\beta_1)\sqrt{n}  \frac{(g\sqrt{\eta}\chi\beta_1-\beta_2)^{n-1}}{\sqrt{(n-1)!}}\ket{n}_B 
\end{multline}

The state then passes through a lossy channel of transmission \(\eta\). Here the loss mode is mode \(C\). 
\begin{equation}
\hat{U}_{BS} \left[ \ket{n}_B\ket{0}_C \right]  = \sum_{k =0}^n \sqrt{\binom{n}{k}} \eta^{k/2} (1-\eta)^{(n-k)/2} \ket{k}_B \ket{n-k}_C 
\end{equation}

\begin{multline}
\ket{\psi(\beta_1, \beta_2)} =\frac{1-\chi^2}{\pi} \frac{1}{\sqrt{1+g_1^2}}  e^{g\sqrt{\eta}\chi(\beta_1\beta_2^*-\beta_1^*\beta_2)/2}  e^{\frac{1}{2}|\alpha-\beta_1|^2(\chi^2-1-\eta\chi^2)} e^{-|g\sqrt{\eta}\chi\beta_1-\beta_2|^2/2}
\\
\sum_{n=0}^\infty \chi^n  (1+g_1\sqrt{\eta}\chi(\alpha-\beta_1)(-g\sqrt{\eta}\chi\beta_1^*+\beta_2^*))\frac{(g\sqrt{\eta}\chi\beta_1-\beta_2)^n}{\sqrt{n!}}\sum_{k =0}^n \sqrt{\binom{n}{k}} \eta^{k/2} (1-\eta)^{(n-k)/2} \ket{k}_B \ket{n-k}_C 
\\
+\sum_{n=1}^\infty \chi^n g_1\sqrt{\eta}\chi(\alpha-\beta_1)\sqrt{n}  \frac{(g\sqrt{\eta}\chi\beta_1-\beta_2)^{n-1}}{\sqrt{(n-1)!}}\sum_{k =0}^n \sqrt{\binom{n}{k}} \eta^{k/2} (1-\eta)^{(n-k)/2} \ket{k}_B \ket{n-k}_C 
\end{multline}
Truncate to order 1 in \(\chi\), this is a good approximation as long as \(\chi\) is kept small (\(\chi\ll 1\)):

\begin{multline}
\ket{\psi(\beta_1, \beta_2)} =\frac{1-\chi^2}{\pi} \frac{1}{\sqrt{1+g_1^2}}  e^{g\sqrt{\eta}\chi(\beta_1\beta_2^*-\beta_1^*\beta_2)/2}  e^{\frac{1}{2}|\alpha-\beta_1|^2(\chi^2-1-\eta\chi^2)} e^{-|g\sqrt{\eta}\chi\beta_1-\beta_2|^2/2}
\\
 \bigg( \left(1+g_1\sqrt{\eta}\chi\left(\alpha-\beta_1\right)\left(-g\sqrt{\eta}\chi\beta_1^*+\beta_2^*\right)\right)  \ket{0}_B \ket{0}_C 
\\
+\chi \left( g_1\sqrt{\eta}\chi\left(\alpha-\beta_1\right) + \left(1+g_1\sqrt{\eta}\chi\left( \alpha-\beta_1 \right) \left( -g\sqrt{\eta}\chi\beta_1^*+\beta_2^* \right) \right) \left( g\sqrt{\eta}\chi\beta_1-\beta_2 \right)   \right) \left(   \left( 1-\eta \right)^{1/2} \ket{0}_B \ket{1}_C + \eta^{1/2}  \ket{1}_B \ket{0}_C \right) \bigg)
 \end{multline}
We then act an NLA on the state (mode \(B\)), with gain \(g_2\):
\begin{multline}
\ket{\psi(\beta_1, \beta_2)} =\frac{1-\chi^2}{\pi} \frac{1}{\sqrt{1+g_1^2}}  \frac{1}{\sqrt{1+g_2^2}}  e^{g\sqrt{\eta}\chi(\beta_1\beta_2^*-\beta_1^*\beta_2)/2}  e^{\frac{1}{2}|\alpha-\beta_1|^2(\chi^2-1-\eta\chi^2)} e^{-|g\sqrt{\eta}\chi\beta_1-\beta_2|^2/2}
\\
 \bigg( \left(1+g_1\sqrt{\eta}\chi\left(\alpha-\beta_1\right)\left(-g\sqrt{\eta}\chi\beta_1^*+\beta_2^*\right)\right)  \ket{0}_B \ket{0}_C 
\\
+\chi \left( g_1\sqrt{\eta}\chi\left(\alpha-\beta_1\right) + \left(1+g_1\sqrt{\eta}\chi\left( \alpha-\beta_1 \right) \left( -g\sqrt{\eta}\chi\beta_1^*+\beta_2^* \right) \right) \left( g\sqrt{\eta}\chi\beta_1-\beta_2 \right)   \right) \left(  \sqrt{1-\eta} \ket{0}_B \ket{1}_C +g_2 \sqrt{\eta} \ket{1}_B \ket{0}_C \right) \bigg)
 \end{multline}
 We then perform a displacement by the measurement result \(\beta_2\) scaled by the gain of the NLA \(g_2\), the strength of entanglement \(\chi\) and transmission of the channel \(\eta\). 
The output state (un-normalised) after two iterations of the protocol is:
\begin{multline}
\ket{\psi(\beta_1, \beta_2)} =\frac{1-\chi^2}{\pi} \frac{1}{\sqrt{1+g_1^2}}  \frac{1}{\sqrt{1+g_2^2}}  e^{g_1 \sqrt{\eta}\chi(\beta_1\beta_2^*-\beta_1^*\beta_2)/2}  e^{\frac{1}{2}|\alpha-\beta_1|^2(\chi^2-1-\eta\chi^2)} e^{-|g_1\sqrt{\eta}\chi\beta_1-\beta_2|^2/2}
\\
\hat{D}_B(g_2\sqrt{\eta}\chi\beta_2) \bigg( \left(1+g_1\sqrt{\eta}\chi\left(\alpha-\beta_1\right)\left(-g_1 \sqrt{\eta}\chi\beta_1^*+\beta_2^*\right)\right)  \ket{0}_B \ket{0}_C 
\\
+\chi \left( g_1\sqrt{\eta}\chi\left(\alpha-\beta_1\right) + \left(1+g_1\sqrt{\eta}\chi\left( \alpha-\beta_1 \right) \left( -g_1 \sqrt{\eta}\chi\beta_1^*+\beta_2^* \right) \right) \left( g_1 \sqrt{\eta}\chi\beta_1-\beta_2 \right)   \right) \left(  \sqrt{1-\eta} \ket{0}_B \ket{1}_C +g_2 \sqrt{\eta} \ket{1}_B \ket{0}_C \right) \bigg)
\label{eq:2out}
 \end{multline}
 
 We now proceed to derive the output state of the complete concatenated protocol in Fig.~\ref{fig:Concatenate2}. This involves taking the state after the first dual homodyne measurement, feeding it into two iterations of the error correction protocol and then acting an NLA and displacement. Results will also have to be averaged over the three complex valued measurement outcomes \(\beta_1\), \(\beta_2\) and \(\beta_3\).

We begin with the state after the first dual homodyne measurement where \(\beta_3\) is the measurement outcome:
 \begin{equation}
 \ket{\psi(\beta_3)} =\sqrt{\frac{1-\chi_3^2}{\pi}} e^{\frac{1}{2}|\alpha-\beta_3|^2(\chi_3^2-1)}\ket{\chi_3\left(\alpha-\beta_3\right)}
 \label{eq:concIn}
 \end{equation}
 
 which is a coherent state with amplitude \(\chi_3\left(\alpha-\beta_3\right)\). 
With input \(\ket{\alpha}\), the output state (un-normalised) after two  iterations of the protocol is \eqref{eq:2out}. With input state \eqref{eq:concIn}, the state after two iterations of the protocol is: 
 
 \begin{multline}
 \ket{\psi(\chi\left(\alpha-\beta_3\right), \beta_1, \beta_2)} =
 \sqrt{\frac{1-\chi_3^2}{\pi}} e^{\frac{1}{2}|\alpha-\beta_3|^2(\chi_3^2-1)} 
 \\
 \frac{1-\chi^2}{\pi} \frac{1}{\sqrt{1+g_1^2}}  \frac{1}{\sqrt{1+g_2^2}}  e^{g \sqrt{\eta}\chi(\beta_1\beta_2^*-\beta_1^*\beta_2)/2}  e^{\frac{1}{2}|\chi_3\left(\alpha-\beta_3\right)-\beta_1|^2(\chi^2-1-\eta\chi^2)} e^{-|g\sqrt{\eta}\chi\beta_1-\beta_2|^2/2}
\\
\hat{D}_B(g_2\sqrt{\eta}\chi\beta_2) 
\bigg[
 \left(1+g_1\sqrt{\eta}\chi\left(\chi_3\left(\alpha-\beta_3\right)-\beta_1\right)\left(-g_1 \sqrt{\eta}\chi\beta_1^*+\beta_2^*\right)\right)  \ket{0}_B \ket{0}_C 
\\
+\chi \left( g_1\sqrt{\eta}\chi\left(\chi_3\left(\alpha-\beta_3\right)-\beta_1\right) + \left(1+g_1\sqrt{\eta}\chi\left(\chi_3\left(\alpha-\beta_3\right)-\beta_1 \right) \left( -g_1 \sqrt{\eta}\chi\beta_1^*+\beta_2^* \right) \right) \left( g_1 \sqrt{\eta}\chi\beta_1-\beta_2 \right)   \right) 
\\
\left(  \sqrt{1-\eta} \ket{0}_B \ket{1}_C +g_2 \sqrt{\eta} \ket{1}_B \ket{0}_C \right) 
\bigg]
 \end{multline}

We define the following variables:
\begin{align}
\kappa &=  \left(1+g_1\sqrt{\eta}\chi\left(\chi_3\left(\alpha-\beta_3\right)-\beta_1\right)\left(-g_1 \sqrt{\eta}\chi\beta_1^*+\beta_2^*\right)\right) 
\\
\lambda &= \chi \left( g_1\sqrt{\eta}\chi\left(\chi_3\left(\alpha-\beta_3\right)-\beta_1\right) + \left(1+g_1\sqrt{\eta}\chi\left( \chi_3\left(\alpha-\beta_3\right)-\beta_1 \right) \left( -g_1 \sqrt{\eta}\chi\beta_1^*+\beta_2^* \right) \right) \left( g_1 \sqrt{\eta}\chi\beta_1-\beta_2 \right)   \right) 
\\
C &= \sqrt{\frac{1-\chi_3^2}{\pi}} e^{\frac{1}{2}|\alpha-\beta_3|^2(\chi_3^2-1)} 
 \frac{1-\chi^2}{\pi} \frac{1}{\sqrt{1+g_1^2}}  \frac{1}{\sqrt{1+g_2^2}}  e^{g \sqrt{\eta}\chi(\beta_1\beta_2^*-\beta_1^*\beta_2)/2}  e^{\frac{1}{2}|\chi_3\left(\alpha-\beta_3\right)-\beta_1|^2(\chi^2-1-\eta\chi^2)} e^{-|g\sqrt{\eta}\chi\beta_1-\beta_2|^2/2}
\end{align}
This simplifies the output state which can now be written as:
\begin{equation}
\ket{\psi(\beta_3, \beta_1, \beta_2)} = C 
\hat{D}_B(g_2\sqrt{\eta}\chi\beta_2) 
\bigg[
\kappa \ket{0}_B \ket{0}_C +\lambda \left(  \sqrt{1-\eta} \ket{0}_B \ket{1}_C +g_2 \sqrt{\eta} \ket{1}_B \ket{0}_C \right) 
\bigg]
\end{equation}
Acting the displacement operator \(\hat{D}_B(g_2\sqrt{\eta}\chi\beta_2) \) on the photon number states:
\begin{equation}
\hat{D}_B(g_2\sqrt{\eta}\chi\beta_2)  \ket{0}_B  = \ket{g_2\sqrt{\eta}\chi\beta_2}_B = e^{-|g_2\sqrt{\eta}\chi\beta_2|^2/2} \left( \ket{0}_B + g_2\sqrt{\eta}\chi\beta_2 \ket{1}_B+... \right) 
\end{equation}
\begin{equation}
\hat{D}_B(g_2\sqrt{\eta}\chi\beta_2)  \ket{1}_B  =  e^{-|g_2\sqrt{\eta}\chi\beta_2|^2/2} \left( -g_2\sqrt{\eta}\chi\beta_2^* \ket{0}_B +\left(1 - |g_2\sqrt{\eta}\chi\beta_2|^2\right) \ket{1}_B+... \right) 
\end{equation}

\begin{multline}
\ket{\psi(\beta_3, \beta_1, \beta_2)} = C e^{-|g_2\sqrt{\eta}\chi\beta_2|^2 /2} 
\bigg[
 \left(\kappa \ket{0}_B \ket{0}_C + \kappa g_2\sqrt{\eta}\chi\beta_2 \ket{1}_B \ket{0}_C +... \right)   \\
 +   \left(  \lambda\sqrt{1-\eta} \ket{0}_B  \ket{1}_C +   \lambda\sqrt{1-\eta} g_2\sqrt{\eta}\chi\beta_2 \ket{1}_B  \ket{1}_C+... \right)  
\\
+ \left( - \lambda g_2 \sqrt{\eta} g_2\sqrt{\eta}\chi\beta_2^* \ket{0}_B \ket{0}_C   + \lambda g_2 \sqrt{\eta} \left(1 - |g_2\sqrt{\eta}\chi\beta_2|^2\right) \ket{1}_B \ket{0}_C   +... \right)  
\bigg]
\end{multline}
Act 3rd NLA with gain \(g_3\):
\begin{multline}
\ket{\psi(\beta_3, \beta_1, \beta_2)} = \frac{1}{\sqrt{g_3^2+1}} C e^{-|g_2\sqrt{\eta}\chi\beta_2|^2 /2} 
\bigg[
 \left(\kappa \ket{0}_B \ket{0}_C + \kappa g_2\sqrt{\eta}\chi\beta_2 g_3 \ket{1}_B \ket{0}_C +... \right)   \\
 +   \left(  \lambda\sqrt{1-\eta} \ket{0}_B  \ket{1}_C +   \lambda\sqrt{1-\eta} g_2\sqrt{\eta}\chi\beta_2 g_3 \ket{1}_B  \ket{1}_C+... \right)  
\\
+ \left( - \lambda g_2 \sqrt{\eta} g_2\sqrt{\eta}\chi\beta_2^* \ket{0}_B \ket{0}_C   + \lambda g_2 \sqrt{\eta} \left(1 - |g_2\sqrt{\eta}\chi\beta_2|^2\right) g_3  \ket{1}_B \ket{0}_C   +... \right)  
\bigg]
\end{multline}
Displace state by \(\hat{D}_B(g_3\sqrt{\eta}\chi\beta_3) \) to give the final output state of the entire concatenated protocol shown in Fig.~\ref{fig:Concatenate2}:
\begin{multline}
\ket{\psi(\beta_3, \beta_1, \beta_2)} = \frac{1}{\sqrt{g_3^2+1}} C e^{-|g_2\sqrt{\eta}\chi\beta_2|^2 /2}  \hat{D}_B(g_3\sqrt{\eta}\chi_3\beta_3)
\bigg[
 \left( \kappa  - \lambda g_2 \sqrt{\eta} g_2\sqrt{\eta}\chi\beta_2^* \right)\ket{0}_B \ket{0}_C   
 \\
 +\left(  \kappa g_2\sqrt{\eta}\chi\beta_2 g_3 + \lambda g_2 g_3  \sqrt{\eta} \left(1 - |g_2 \sqrt{\eta}\chi\beta_2|^2\right)  \right)\ket{1}_B \ket{0}_C  
 +   \lambda\sqrt{1-\eta} \ket{0}_B  \ket{1}_C +   \lambda\sqrt{1-\eta} g_2\sqrt{\eta}\chi\beta_2 g_3 \ket{1}_B  \ket{1}_C 
\bigg]
\label{eq:ConcOut}
\end{multline}

The success probability and variance were calculated following the same formula as \eqref{eq:Pformula} and \eqref{Vformula}, this time using the output state \eqref{eq:ConcOut}. Numerical integration was performed to average the result over \(\beta_1\), \(\beta_2\) and \(\beta_3\).

\end{widetext}

\bibliographystyle{apsrev}
\bibliography{RefList}

\end{document}